\documentclass[conference]{IEEEtran}
\ifCLASSINFOpdf
\else
\fi

\usepackage{graphicx}
\usepackage[colorinlistoftodos]{todonotes}
\usepackage{amsmath}
\usepackage{multicol}			
\usepackage{multirow}
\usepackage{hhline}
\usepackage{longtable}
\usepackage{tabularx}
\usepackage{fancyhdr}			
\usepackage{hyperref} 			
\usepackage{epstopdf}
\DeclareGraphicsExtensions{.eps,.ps,.pdf,.png,.jpg}

\makeatletter
\def\ps@IEEEtitlepagestyle{
  \def\@oddfoot{\mycopyrightnotice}
  \def\@evenfoot{}
}
\def\mycopyrightnotice{
  {\footnotesize
  \begin{minipage}{\textwidth}
  \centering
 ~\copyright~2017 IEEE. Personal use of this material is permitted. Permission from IEEE must be  obtained for all other uses, in any current or future media, \\ including reprinting/republishing this material for advertising or promotional purposes, creating new  collective works, for resale  \\ or redistribution to servers or lists, or reuse of any copyrighted component of this work in other works.\\ DOI:  \href{http://ieeexplore.ieee.org/document/7962714/}{10.1109/ICCW.2017.7962714}
\end{minipage}
  }
}

\begin{document}

%
\title{Optimized LTE Data Transmission Procedures for IoT: Device Side Energy Consumption Analysis}

\author{\IEEEauthorblockN{Pilar Andres-Maldonado, Pablo Ameigeiras, Jonathan Prados-Garzon, Juan J. Ramos-Munoz\\ and Juan M. Lopez-Soler}
\IEEEauthorblockA{Department of Signal Theory, Telematics, and Communications \\ University of Granada \\
Granada, Spain\\
Email: pam91@correo.ugr.es, \{pameigeiras, jpg, jjramos, juanma\}@ugr.es
}}


%


\maketitle


\begin{abstract}
The efficient deployment of Internet of Things (IoT) over cellular networks, such as Long Term Evolution (LTE) or the next generation 5G, entails several challenges. For massive IoT, reducing the energy consumption on the device side becomes essential. One of the main characteristics of massive IoT is small data transmissions. To improve the support of them, the 3GPP has included two novel optimizations in LTE: 
one of them based on the Control Plane (CP), and the other on the User Plane (UP). In this paper, we analyze the average energy consumption per data packet using these two optimizations compared to conventional LTE Service Request procedure. We propose an analytical model to calculate the energy consumption for each procedure based on a Markov chain. 
In the considered scenario, for large and small Inter-Arrival Times (IATs), the results of the three procedures are similar. While for medium IATs CP reduces the energy consumption per packet up to 87\% due to its connection release optimization.
\end{abstract}


%
\IEEEpeerreviewmaketitle

\section{Introduction}
\label{sec:intro}

 


Machine Type Communications (MTC), and more generally the Internet of Things (IoT), are big umbrellas that include a plethora of different applications. 
The Third Generation Partnership Project (3GPP) has classified IoT into two scenarios: massive IoT (mIoT) and ultra-reliable and low latency communications (uRLLC) \cite{Hoymann16}. Both frameworks are identified as use cases either for Long Term Evolution (LTE) and for 5G. The foreseen huge number of connected IoT devices in mIoT, even considering the expected small volume of data per session, will lead to a significant network 
signaling overload. 

Recently, MTC and Narrowband IoT (NB-IoT) 3GPP's tracks have achieved improvements regarding to device complexity, coverage, and power consumption \cite{Hoymann16}. 
In this sense, the 3GPP has included two new mechanisms in LTE, the Control Plane Cellular IoT (CP) optimization and the User Plane Cellular IoT (UP) optimization. The purpose of these procedures is to optimize the data transmission for IoT devices. Hence, we pursue to find out their advantages.

In this paper, we concentrate on the energy consumption of the devices, which is one of the most relevant Key Performance Indicators (KPI) for IoT. Specifically, we answer the following questions: i) What is the energy reduction achieved by IoT devices with these new mechanisms, and ii) Which devices benefit from using these mechanisms, as we expect different Inter-Arrival Times (IATs) for different devices.

To answer these questions, we will use the conventional data transmission through Service Request (SR), 
with an inactivity timer ($T_{i}$) of 10s, as the baseline procedure.

There are other works focused on energy consumption analysis for LTE. The authors of \cite{Wang16} analyze the impact of the Discontinuous Reception (DRX) mechanism. 3GPP's technical report \cite{3gpp45820} includes the battery lifetime estimation during an uplink transmission for different cellular IoT deployments. However, to the best of our knowledge, there are no energy consumption comparisons at the IoT device side between the new 3GPP IoT optimizations and the baseline SR scheme.



For this goal, we study the signaling, resources allocated, and steps performed in each procedure. Then, we estimate and compare the average energy consumption per packet and device's battery lifetime for each scheme. This work extends the analytical model of \cite{Madueño16}. From the radio access capacity analysis of \cite{Madueño16}, we add the analysis of the energy consumption per data packet, and the power saving mechanisms. In the considered scenario, the results show that for large and small IATs, the performance of the three procedures is similar. However, for medium IATs the CP optimization reduces up to 87\% the energy consumption per packet due to its connection release optimization.

The remainder of this paper is organized as follows. Section \ref{sec:LTEIoT} gives an introduction to LTE data transmission, and power saving mechanisms. Section \ref{sec:analysis} provides the analytical model and the energy consumption estimation. Section \ref{sec:numresults} shows the numerical results. Finally, Section \ref{sec:conclusion} sums up the conclusions.

\section{LTE Optimizations for IoT}
\label{sec:LTEIoT}

In this section, we first explain the data transmission scheme used in LTE and the two data transmission optimizations added 
for IoT. 
Then, we describe two LTE power saving mechanisms.

\subsection{Data Transmission Procedures for IoT}
\label{subsec:SR}

In LTE, the transmission of data packets requires an established Radio Resource Control (RRC) connection between the device and the evolved NodeB (eNB). When an RRC connection has been established, the device is in \texttt{RRC Connected}. After a device's inactivity period of $T_{i}$, the RRC connection is released through an S1-Release procedure, then the device changes to \texttt{RRC Idle}.

If an \texttt{RRC Idle} device wants to communicate with the network, it has to perform the SR to establish the RRC connection and request resources again. Part of the work of the 3GPP for IoT focus on reducing the signaling overhead required to establish and release a communication channel compared to conventional LTE. There are two solutions introduced \cite{3gpp23401}: one of them based on the Control Plane (CP), and the other on the User Plane (UP). Therefore, the procedures available to send data packets from an \texttt{RRC Idle} IoT device in LTE are:



\subsubsection{Service Request (SR) Procedure}

\begin{figure}[!bt] 
	\centering
    \includegraphics[width=0.92\columnwidth,height=6.8cm]{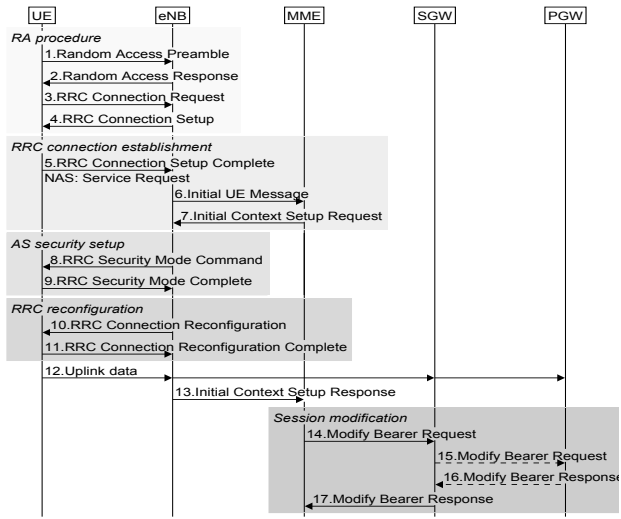}
	\caption{User Equipment (UE) triggered Service Request procedure \cite{3gpp23401}.}
	\label{fig:SR}
\end{figure}

This is the conventional data transmission procedure. Figure \ref{fig:SR} shows a typical sequence of signaling messages. 
The first four messages comprise the contention-based random access (RA) procedure. Steps 3 and 4 of the RA procedure are used as part of the RRC connection establishment. The subsequent signaling messages perform: device's authentication in the Mobility Management Entity (MME) through Non Access Stratum (NAS) security level (steps 6 and 7), Access Stratum (AS) security context establishment between the device and the eNB (steps 8 and 9), RRC reconfiguration (steps 10 and 11), and data bearers' establishment with other core entities (steps 14 to 17). 

\subsubsection{Control Plane Cellular IoT Optimization (CP)}

\begin{figure}[!bt] 
	\centering
    \includegraphics[width=0.92\columnwidth,height=4.5cm]{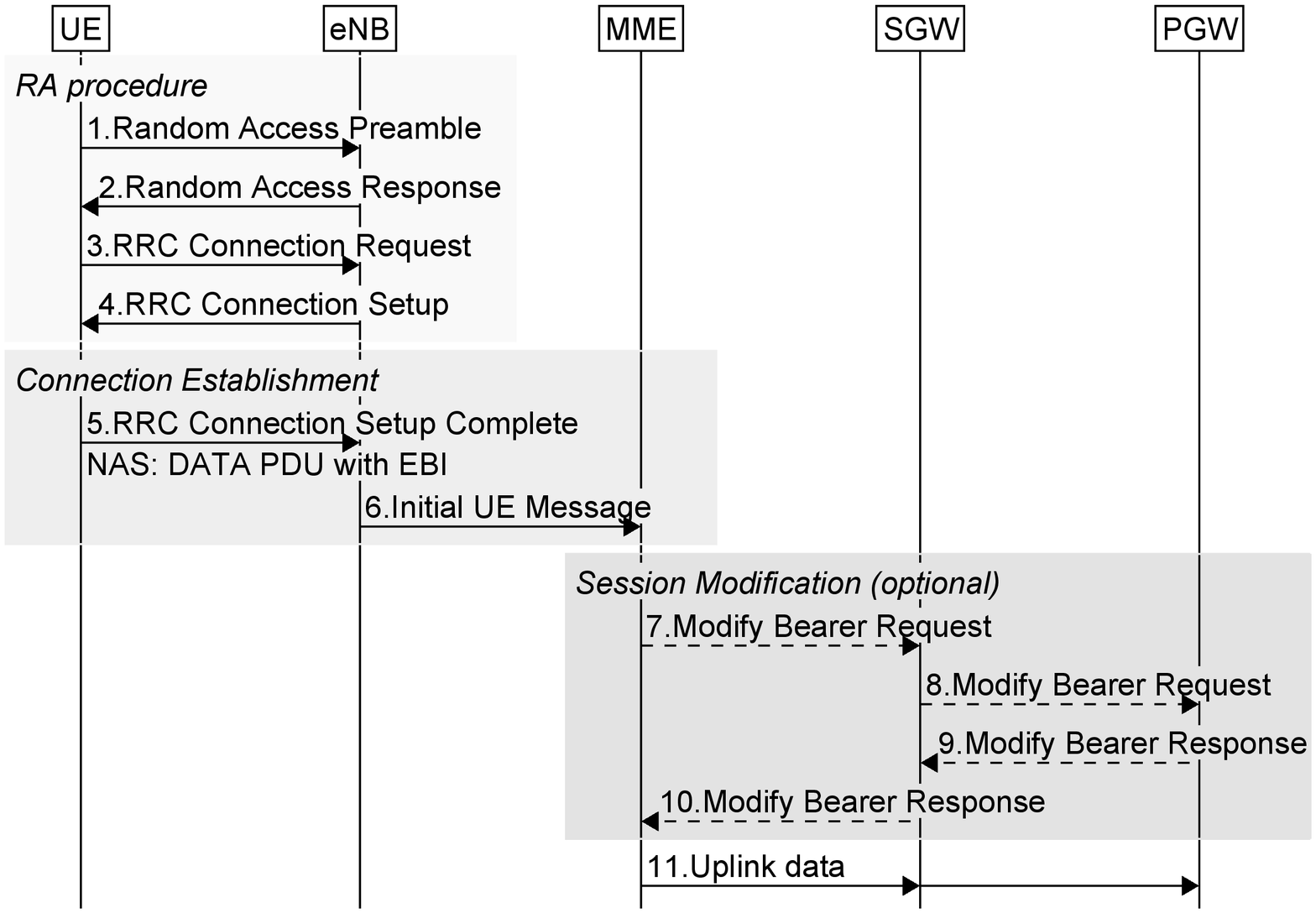}
	\caption{Mobile Originated data transport in control plane solution \cite{3gpp23401}.}
	\label{fig:CP}
\end{figure}

This optimization uses the control plane to forward the device's data packets. To do that, they are sent through NAS messages to the MME. Figure \ref{fig:CP} shows a typical sequence of signaling messages. Contrary to conventional SR, the MME is responsible for the data packets' security check, and forwarding them to the Serving Gateway (SGW) through the new S11-U bearer.

This procedure does not apply AS security, and there is no RRC reconfiguration, which represents a reduction of signaling messages over the radio interface. Furthermore, the device can include in the NAS message the Release Assistance Indication to the MME. This information can notify whether no further data transmission is expected. 
Then, the MME could release the connection immediately if there is no pending traffic, and S1-U bearer (between eNB and SGW) is not established. 

\subsubsection{User Plane Cellular IoT Optimization (UP)}
 
\begin{figure}[!bt] 
	\centering
    \includegraphics[width=0.85\columnwidth,height=4.5cm]{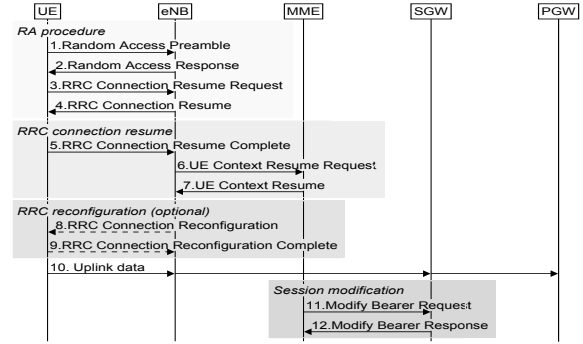}
	\caption{UE initiated Connection Resume procedure in user plane solution \cite{3gpp23401}.}
	\label{fig:UP}
\end{figure}

The transmission of data in this optimization is over the data plane, by means of preserving the RRC context instead of release it. 
Two control procedures are defined: Connection Resume and Suspend. Both are similar to two conventional control procedures, SR and S1-Release, respectively. However, the signaling messages slightly change. Compared with SR, the context of the device is stored in the device and the network. Then, there is no need to reestablish the AS security context. 
Figure \ref{fig:UP} shows a typical sequence of signaling messages.

\subsection{Power Saving Functions}
\label{subsec:power}

While the device has an RRC connection established, it has to listen to the network in order to decode Physical Downlink Control Channel (PDCCH) in every subframe. 
For a device, it implies to keep the radio receptor always on. Therefore, it is a waste of energy when there is not data available. 
To improve the device's battery lifetime, the 3GPP has introduced enhancements to enable devices to communicate on a \textit{per-need} basis, increasing the time in low-power consumption mode.  

\subsubsection{Discontinuous Reception (DRX)}
\label{subsubsec:DRX}

Is a mechanism that allows the device to stop monitoring radio channels and to enter low-power consumption mode for short periods of time. 

In LTE, this functionality has two modes, 
called \texttt{Connected Mode DRX} and \texttt{Idle Mode DRX}. Both DRX modes allow the device to monitor the PDCCH channel discontinuously, in order to check if there is data available in \texttt{Connected Mode DRX}, or paging messages in \texttt{Idle Mode DRX}. There are a set of user specific DRX parameters configured based on the device activity. 
For more information about DRX, 
see \cite{3gpp36321}.

\subsubsection{Power Saving Mode (PSM)}
\label{subsubsec:PSM}

Is a device's mode which implies that the device is not reachable, but it is still registered with the network \cite{3gpp23682}. The energy consumption in PSM is almost as being in the power-off state, that is, lower than the energy consumption of \texttt{RRC Idle}. The PSM mode starts after a period of time in which the device is in \texttt{RRC Idle} without further activity, called \textit{Active Time}. Then, the device will enter PSM mode. A device is in PSM until a mobile originated event requires the communication with the network. 
For more information about PSM, see \cite{3gpp23682}.

\section{Energy Consumption Analysis}
\label{sec:analysis}

In this work, we provide a Markov chain analysis of the average energy consumption needed to transfer one data packet for the three control procedures explained before. The analysis is divided into two different parts. First, we model with a Markov chain the behavior of a device. 
Then, we estimate the average energy consumption per packet as a function of the stationary probabilities and the average energy consumption required to complete each event of the Markov chain.

\subsection{Transmission Model}
\label{subsec:transmod}

The basic time unit in the analysis is one LTE subframe (1ms). 
We consider the resources needed for every packet transmission are allocated to the device in units of Resource Block (RB) pairs. 
We assume that the packet generation process of each IoT device follows a Poisson model with rate $\lambda_{app}$ packets per ms. The data rate of the IoT device is derived from its average IAT in ms, therefore $\lambda_{app} = \frac{1}{IAT}$. 

\begin{figure}[!bt] 
	\centering
    \includegraphics[width=\columnwidth,height=2.7cm]{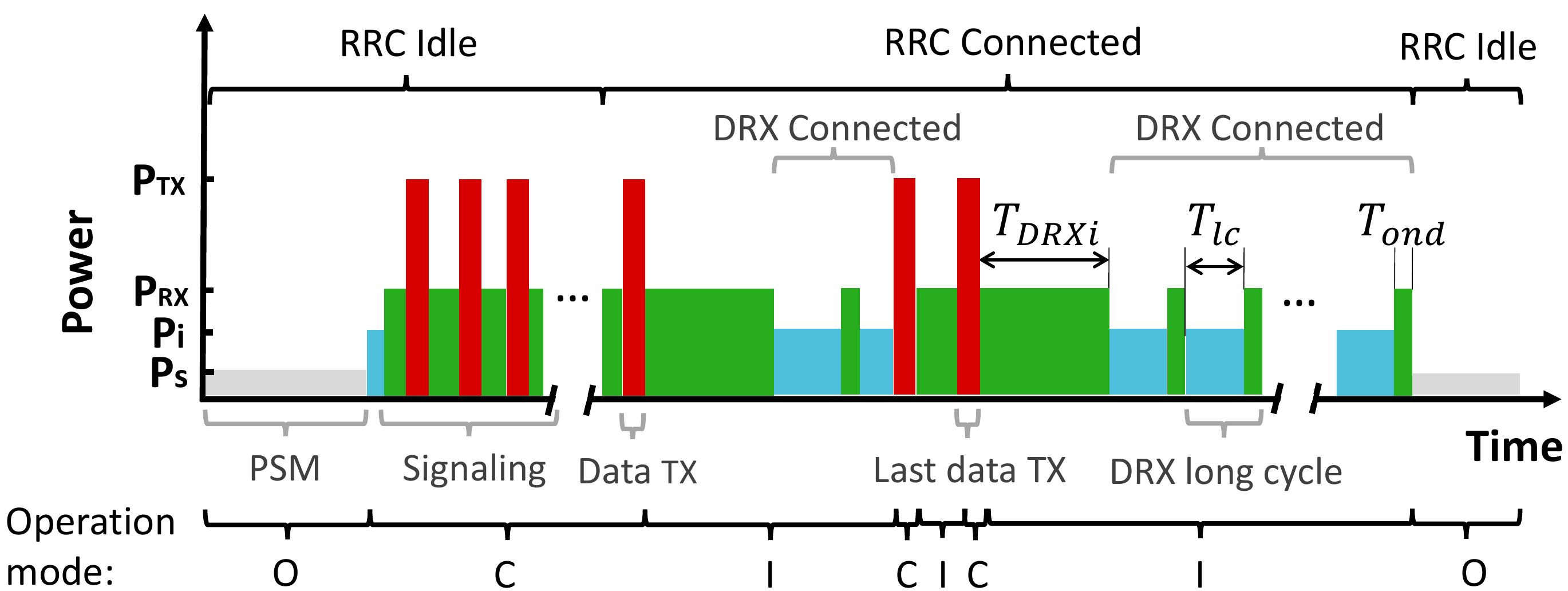}
	\caption{IoT device behavior assumed (O: \textit{Off}, C: \textit{Communication}, I: \textit{Inactive}).}
	\label{fig:potdiag}
\end{figure}

To model the device's behavior, we consider three 
operation modes (see Figure \ref{fig:potdiag}):
\subsubsection{Off} 
The device is in \texttt{RRC Idle}, and uses PSM mechanism to save energy.
\subsubsection{Communication} This operation mode starts when the \texttt{RRC Idle} device wants to send a data packet. It 
involves the steps incurred 
since the device starts its 
RA procedure 
until the device is in \texttt{RRC Connected}. 
This operation mode also includes the subsequent data transmissions while the device is still in \texttt{RRC Connected}. Between transmissions, the device is in \textit{Inactive} operation mode. The last transmission happens when the device is not going to send any data packet within the inactivity timer $T_{i}$. Therefore, this operation mode includes signaling and data packets between the network and the device. 

\subsubsection{Inactive} This operation mode includes the steps when the device is in \texttt{RRC Connected}, but it is not sending data packets. We assume the device uses \texttt{Connected Mode DRX} to save energy. After DRX inactivity time $T_{DRXi}$, we only assume long DRX cycles. These cycles are composed by two timers: $T_{lc}$, which denotes the period of time the device is idle within the DRX cycle, and $T_{ond}$, defined as the period of time the device is listening to the network within the DRX cycle. After the last transmission and its inactive time, the device changes to \texttt{RRC Idle}. 
We assume that whenever the device reaches \texttt{RRC Idle}, it directly changes to \textit{Off} operation mode.
\subsection{Markov Chain Analysis}
\label{subsec:markov}
\begin{figure*} 
	\centering
	\includegraphics[width=0.9\textwidth,height=3.4cm]{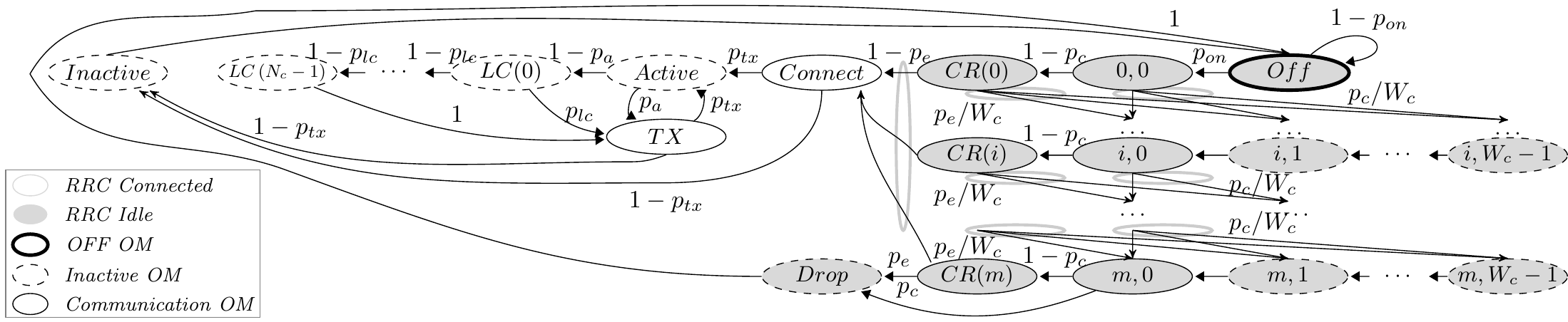}
	\caption{Markov chain model for $m$ retransmissions of a device during \textit{Off}, \textit{Communication} and \textit{Inactive} operation modes (OM) for the three procedures.}
	\label{fig:chain}
\end{figure*}

Figure \ref{fig:chain} depicts the proposed Markov chain to model the considered operation modes. Our model is an extension of the one proposed in \cite{Madueño16}. 

To connect to the network, up to $m$ retransmissions of the RA procedure are allowed. If a device fails to request a connection, the device backs off. The states $\left \{ i,0 \right \}$ represent the $i$th RA attempt. The backoff time is uniformly chosen in the range $\left (0,W_{c}-1 \right )$, being $W_{c}$ the maximum backoff window size. Then, the states $\left \{i,k\right \}$ represent the $k$th backoff counter of the $i$th retransmission. Additionally, $CR\left ( i \right )$ state corresponds to the $i$th connect request attempt, which represents the remainder steps the device has to perform 
to change to \texttt{RRC Connected}, i.e. RRC establishment, AS security reestablishment, RRC reconfiguration, or bearers establishment. The $TX$ state represents the transmission of a data packet while the device is in \texttt{RRC Connected}. Lastly, $Active$ and $LC\left ( n \right )$ respectively represent the period of time the \texttt{RRC Connected} device waits before starts \texttt{Connected Mode DRX}, and the $n$th long DRX cycle. The number of long DRX cycles is denoted by $N_{c}$, derived as $N_{c}=\left \lfloor \frac{T_{i}-T_{DRXi}}{T_{lc}+T_{ond}}\right \rfloor$.

Let $p_{on}$ denote the probability of having uplink traffic in a subframe, expressed as $p_{on} = 1 - e^{-\lambda_{app}}$. Let $p_{c}$ and $p_{e}$ respectively be the probabilities of collision at RA procedure and error at the connection request.
Expressions for $p_{c}$ and $p_{e}$ can be found in \cite{Madueño16}. Let, 
\begin{itemize}
\item $P_{out}$ denote the outage probability, derived as $P_{out}=1-\left( 1-p_{e} \right) \left( 1-p_{c} \right)$.
\item $p_{tx}$ be the data transmission probability before $T_{i}$ expires, expressed as $p_{tx} = 1 - e^{-\lambda_{app} T_{i}}$.
\item  $p_a$ denote the data transmission probability before $T_{DRXi}$ expires, derived as $p_{a} = 1 - e^{-\lambda_{app} T_{DRXi}}$.
\item $p_{lc} = 1 - e^{-\lambda_{app} \left( T_{ls} + T_{ond} \right) }$ be the probability of transmission before $T_{ls} + T_{ond}$ expires.
\end{itemize}

Then, based on \cite{Madueño16}, the transition probabilities of the Markov chain can be expressed as: 

\begin{equation*}
  \label{eq:transitionprob}
  \begin{aligned}
    P\left ( 0,0|Off \right ) &= p_{on}
	\\
	P\left ( CR\left (i  \right )|i,0 \right ) &= 1-p_{c}, \quad \quad \quad \quad \quad \quad \quad i \in \left [ 0, m \right ]
	\\
	P\left ( Connect|CR\left (i  \right ) \right ) &= 1-p_{e}, \quad \quad \quad \quad \quad \quad \quad i \in \left [ 0, m \right ]
	\\
    P\left ( i,k|i-1,0 \right ) &= \frac{p_{c}}{W_{c}}, \quad k \in \left [ 0, W_{c}-1 \right ], i \in \left [ 1, m \right ] 
    \\
    P\left ( i,k|CR\left (i - 1  \right ) \right ) &= \frac{p_{e}}{W_{c}}, \quad k \in \left [ 0, W_{c}-1 \right ], i \in \left [ 1, m \right ]
    \\
    P\left ( Active|Connect \right ) &= P\left ( Active|TX \right ) = p_{tx}
	\end{aligned}
\end{equation*}
\begin{equation*}
  \label{eq:transitionprob2}
  \begin{aligned}
    P\left ( LC\left (0  \right )|Active \right ) &= 1 - p_{a}
  	\\
    P\left ( LC\left (n  \right )|LC\left (n -1 \right ) \right ) &= 1 - p_{lc}, \quad \quad \quad \quad n \in \left [ 1, N_{c}-1 \right ]
    \\
    P\left ( TX|Active \right ) &= p_{a}
    \\
    P\left ( TX|LC\left (n  \right ) \right ) &= p_{lc}, \quad \quad \quad \quad \quad \quad n \in \left [ 0, N_{c}-2 \right ]
    \\
    P\left ( TX|LC\left (N_{c} -1 \right ) \right ) &= 1
    \\
    P\left ( Inactive|TX \right ) &=  P\left ( Inactive|Connect \right ) =1-p_{tx}
    \\
    P\left ( Drop|m,0 \right ) &= p_{c}
    \\
    P\left ( Drop|CR\left (m  \right ) \right ) &= p_{e}
	\\
    P\left ( Off|Drop \right ) &= P\left ( Off|Inactive \right ) = 1
  \end{aligned}
\end{equation*}
Let $b_{j}$ denote the steady state probability that a device is at $j$ state. Then, they can be derived as \cite{Madueño16}:
\begin{equation*}
  \label{eq:steadyprob}
  \begin{aligned}
	b_{off} &= \left ( 1-p_{on} \right )b_{off} + b_{drop} + b_{inactive}
    \\
	b_{0,0} &= p_{on}b_{off}
    \\
	b_{i,0} &= p_{e}b_{CR\left ( i-1 \right )} + p_{c}b_{ i-1,0 } = \left ( p_{e}\left ( 1-p_{c} \right ) + p_{c} \right )^{i}b_{0,0}
	\\
	b_{i,k} &= \tfrac{W_{c}-k}{W_{c}}b_{i,0} =\tfrac{W_{c}-k}{W_{c}} \left ( p_{e}\left ( 1-p_{c} \right ) + p_{c} \right )^{i} b_{0,0}    
    \\
    b_{CR\left ( i \right )} &= \left ( 1-p_{c} \right )b_{i,0}=\left ( 1-p_{c} \right )\left ( p_{e}\left ( 1-p_{c} \right ) + p_{c} \right )^{i}b_{0,0}
    \\
	b_{drop} &= p_{e}b_{CR\left ( m \right )} + p_{c}b_{ m,0 } = \left ( p_{e}\left ( 1-p_{c} \right ) + p_{c} \right )^{m+1}b_{0,0}
    \\
    b_{connect} &= \sum_{i=0}^{m} \left ( 1-p_{e} \right )b_{CR\left ( i \right )} 
    \\
    b_{active} &= p_{tx}\left ( b_{connect} + b_{TX}\right ) = \frac{p_{tx}}{1-p_{tx}}b_{connect}
    \\
    b_{LC(n)} &= \left ( 1-p_{lc} \right )^n \left ( 1-p_{a} \right )b_{active}
    \\
    b_{TX} &= p_{a}b_{active} + p_{lc}\sum_{n=0}^{N_{c}-2}b_{LC(n)} + b_{LC(N_{c}-1)} = b_{active} 
    \\
    b_{inactive} &= \left ( 1-p_{tx} \right )\left ( b_{TX} + b_{connect}\right ) 
  \end{aligned}
\end{equation*}
By imposing the probability normalization condition:
\begin{equation*}
  \label{eq:normprob}
  \begin{aligned}
1 &= b_{off} + b_{connect} + b_{0,0} + b_{drop} + b_{active} + b_{TX} +
\\
&  b_{inactive} + \sum_{i=1}^{m} \sum_{k=0}^{W_{c}-1} b_{i,k} + \sum_{i=0}^{m} b_{CR\left ( i \right )} + \sum_{n=0}^{N_{c}-1} b_{LC\left ( n \right )}
  \end{aligned}
\end{equation*}
then, we obtain $b_{off}$ as

\begin{equation*}
  \label{eq:normprob2}
  \begin{aligned}
b_{off} &= \left ( 1 + p_{on} \left ( 1 + s^{m+1} + \frac{\left ( 1-s^{m+1} \right )\left ( 1-p_{c} \right )}{1-s} + \right. \right.
\\
&\left.\left. \quad \frac{ s\left ( 1-s^m \right ) \left ( 1+W_{c} \right )}{2\left ( 1-s \right )} + \left ( 1-s^{m+1} \right ) aux \right ) \right )^{-1}
  \end{aligned}
\end{equation*}

where $s= p_{e}\left ( 1-p_{c} \right ) + p_{c}$, and $aux = 2 - p_{tx} + \tfrac{p_{tx}}{1-p_{tx}} \left ( 3 - p_{tx} + \left( 1-p_{a} \right )\tfrac{1-\left ( 1-p_{lc} \right )^{N{c}}}{p_{lc}} \right )$.

\subsection{Estimation of the Average Energy Consumption per Packet}
\label{subsec:energypp}

The average energy consumption per packet can be derived from the power and duration of each state of the Markov chain
\cite{Wang16}. Then, let $E_{j}$ be the average energy consumption of the $j$ state. Let $N_{p}$ denote the number of packets sent while the device is in \texttt{RRC Connected}. The average energy consumption per packet $E_{p}$ can be calculated as:

\begin{equation*}
	E_{p} = \frac{\sum_{j}b_{j}E_{j}}{N_{p}}
\end{equation*}

where $N_{p} = b_{connect}\left ( 1 + \sum_{n=1}^{\infty} {p_{tx}}^{n}\left ( 1-p_{tx} \right )n \right )$. The expression of each $b_{j}$ is already derived, so we need to derive each element of $E_{j}$. For the power analysis, we consider four different device's power levels \cite{Haro15}:
\begin{itemize}
\item Sleep ($P_{s}$): The device consumes minimum power. The low-power clock is on while the device is sleeping.
\item Idle ($P_{i}$): The device is not transmitting or receiving packets. It keeps on the accurate clock to be able to maintain the synchronization with the air interface.
\item RX ($P_{RX}$): The device is expecting a packet from the network, or processing a response to the network. Then, the RX branch of the device is on.
\item TX ($P_{TX}$): The device is transmitting a packet to the network, the TX branch of the device is on. To obtain the transmit power required, we use the 3GPP's power control equations of the different uplink physical channels considered in the analysis (see \cite{3gpp36213}).
\end{itemize}

Next, we derive  $E_{j}$ for each state $j$ of the Markov chain. For these calculations, we use the latency analysis of \cite{3gpp36912}, and the messages sizes of \cite{Madueño16,3gpp37869}.
Table \ref{table:var} shows the notation.

\begin{itemize}
\item $Off$ state: The device does not transmit uplink packet in current subframe $E_{off}^{SR}= E_{off}^{CP}= E_{off}^{UP}= P_{s}$

\item $0,0$ state: The device performs the RA procedure
\begin{equation*}
	E_{0,0}^{SR} = E_{0,0}^{CP} = E_{0,0}^{UP} = T_{pre}P_{i} + T_{RA_{RX}}P_{RX} + P_{pre}
\end{equation*}
\item $i,0$ state: After a unsuccessful connection and a backoff time, the device retries the RA procedure. Then $\quad \quad \quad \quad E_{i,0}^{SR} = E_{i,0}^{CP} = E_{i,0}^{UP} = E_{0,0}^{SR}$

\item $i,k$ state: $k$th backoff wait $E_{i,k}^{SR} = E_{i,k}^{CP} = E_{i,k}^{UP} = P_{i}$

\item $CR$ state: The device performs a connection request. For CP procedure, the data transmission happens at this stage sending the data into the RRC Connection Setup Complete message ($B_{comp_{CP}}$)
\begin{equation*}
	\begin{aligned}
		E_{CR\left ( i \right )}^{SR} &= T_{CR_{RX}}^{SR}P_{RX} + P_{RBp}\left ( \left \lceil \tfrac{B_{req}}{B_{RBp}} \right \rceil + \right.
        \\
         & \quad \quad \quad \quad \quad \left. \left \lceil \tfrac{B_{comp}}{B_{RBp}} \right \rceil + \left \lceil \tfrac{B_{s-comp}}{B_{RBp}} \right \rceil + \left \lceil \tfrac{B_{r-UL}}{B_{RBp}} \right \rceil \right ) 
        \\
        E_{CR\left( i \right )}^{CP} &= T_{CR_{RX}}^{CP}P_{RX} + P_{RBp}\left ( \left \lceil \tfrac{B_{req}}{B_{RBp}}  \right \rceil + \left \lceil \tfrac{B_{comp_{CP}}}{B_{RBp}} \right \rceil \right )
        \\
        E_{CR\left( i \right )}^{UP} &= T_{CR_{RX}}^{UP}P_{RX} + P_{RBp}\left ( \left \lceil \tfrac{B_{req}}{B_{RBp}} \right \rceil + \left \lceil  \tfrac{B_{comp}}{B_{RBp}} \right \rceil\right )
	\end{aligned}
\end{equation*}

\item $Connect$ state: After a successful connection, the device sends its data packet. For CP procedure this packet has been already sent, so $E_{connect}^{CP} = 0$
\begin{equation*}
	\begin{aligned}
    E_{connect}^{SR} &= E_{connect}^{UP} = P_{RBp}\left ( \left \lceil \tfrac{B_{data}}{B_{RBp}} \right \rceil  \right ) 
	\end{aligned}
\end{equation*}

\item $Active$ state: The device listens to the network, no uplink packet until time \textit{l} 
\begin{equation*}
	\begin{aligned}
    E_{active} &=  \left (\sum_{l=1}^{T_{DRXi}-1}e^{-\lambda _{app}\left ( l-1 \right )} \left ( 1- e^{-\lambda _{app}}\right )l + \right.
    \\
    & \left. \quad \quad \quad e^{-\lambda _{app}\left ( T_{DRXi}-1 \right )}T_{DRXi}  \right )P_{RX}
    \\
    E_{active}^{SR} &= E_{active}^{CP} = E_{active}^{UP} =  E_{active}
	\end{aligned}
\end{equation*}

\item $LC$ state: The device is in a long DRX cycle, no uplink packet until time \textit{l}
\begin{equation*}
	\begin{aligned}
    E_{LC} &=  \sum_{l=1}^{ T_{lc} +T_{ond} - 1}e^{-\lambda _{app}\left ( l-1 \right )} \left ( 1- e^{-\lambda _{app}}\right )l P_{i} + 
    \\
    &  \quad \quad \quad e^{-\lambda _{app}\left ( T_{lc} +T_{ond} -1 \right )} \left ( T_{lc}P_{i} +T_{ond}P_{RX} \right )
    \\
    E_{LC}^{SR} &= E_{LC}^{CP} = E_{LC}^{UP} =  E_{LC}
	\end{aligned}
\end{equation*}
\item $TX$ state: The device sends a data packet. For CP procedure, the data packet is sent as NAS signaling into a RRC UL Information Transfer message ($B_{data_{CP}}$)
\begin{equation*}
	\begin{aligned}
    E_{TX}^{SR} &= E_{TX}^{UP} = P_{RBp}\left ( \left \lceil \tfrac{B_{data}}{B_{RBp}} \right \rceil  \right ) 
    \\
    E_{TX}^{CP} &= P_{RBp}\left ( \left \lceil \tfrac{B_{data_{CP}}}{B_{RBp}} \right \rceil  \right ) 
	\end{aligned}
\end{equation*}

\item $Inactive$ state: After the last transmission, the device changes to \textit{Inactive}. The device stays in this state until the connection is released. For CP procedure, we assume the release of the connection is done after a period sufficiently long ($T_{wait})$ such that a waiting data packet can be delivered to the device from the MME \cite{3gpp23401}
\begin{equation*}
	\begin{aligned}
    E_{inactive}^{SR} &= E_{inactive}^{UP} = \left (T_{DRXi} + N_{c}T_{ond} \right )P_{RX} +
    \\
    & \quad \quad \quad \quad \quad \quad \left (N_{c}T_{lc} + T_{spare}  \right )P_{i}
    \\
    E_{inactive}^{CP} &= T_{wait}P_{RX}
	\end{aligned}
\end{equation*}

where $T_{spare} = T_{i} - \left ( T_{DRXi} + N_{c} \left ( T_{lc}+T_{ond} \right ) \right )$ denotes the spare time after DRX and before the expiration of $T_{i}$.

\item $Drop$ state: If the device fails all attempts of the RA procedure and connection requests, the device drops the data packet. There is not energy consumption in this state, then, $E_{drop}^{SR} = E_{drop}^{CP} = E_{drop}^{UP} =0$.

\end{itemize}

\begin{table}[tb]
\centering
\caption{Variable Notation and Values}
\label{table:var}
\begin{tabular}{l|c|l}
\hline
\textbf{Variable} & \textbf{Value}  & \textbf{Description}                               \\ \hline
$E_{j}^{z}$                              & Variable                      & \begin{tabular}[c|]{@{}l@{}}Average energy concumption in state $j$  \\ for the procedure $z \in \left \{SR,CP,UP\right \}$ \end{tabular}                                                             \\
$P_{s}$                                  & 0.03                          & Sleep power consumption (mW) \cite{Haro15}                                                                                                                                                          \\
$P_{i}$                                  & 10                            & Idle power consumption (mW) \cite{Haro15}                                                                                                                                                       \\
$P_{RX}$                                 & 100                           & RX power consumption (mW) \cite{Haro15}                                                                                                                                                             \\
$P_{TX_{MAX}}$                           & 200                           & TX max power consumption (mW) \cite{3gpp36213}                                                                                                                                                  \\
$P_{RBp}$                                & 32.18                          & \begin{tabular}[c|]{@{}l@{}} TX power consumption per RB pair (mW) \\ (700m, $P_{O\_PUSCH}=-100dBm$, $\alpha=1$, \\ $\Delta_{TF}=0dB$, and $f_{c}=0$)  \end{tabular}                      \\
$P_{pre}$                                & 32.18                          & \begin{tabular}[c|]{@{}l@{}} Preamble TX power consumption (mW) \\ (700m, $PreambleInitialRTP=-100dBm$,\\
$\Delta_{pre}=0dB$, and $RampingStep=0dBm$)  \end{tabular}                      \\
$T_{pre}$                                & 2.5                           &  Time before preamble transmission (ms)    \\
$T_{RA_{RX}}$                            & 10                            & RX time to perform RA (ms)                                                                                                                                           \\
$T_{CR_{RX}}^{z}$                        & $41,16,16$ & \begin{tabular}[c|]{@{}l@{}} Sum of processing delays (UE and eNB) to \\ establish a connection (ms), $z$\scalebox{0.8}{$\in \left \{SR,CP,UP\right \}$} \end{tabular} \\
$T_{DRXi}$                               & 200                           & DRX Inactivity timer (ms) \cite{webDRX}                                                                                                                                                                      \\
$T_{ond}$                                & 4                             & \texttt{RRC Connected} On duration timer (ms)\cite{webDRX}                                                                                                                                                           \\
$T_{lc}$                                 & 80                            & DRX Long Cycle (ms) \cite{webDRX}                                                                                                                                                                            \\
$T_i$                                    & 10000                         & Inactivity timer (ms)                                                                                                                                                                                          \\
$T_{wait}$                                 & 54                            & \begin{tabular}[c|]{@{}l@{}}S1 processing and transfer delay (ms) \cite{3gpp083917}\end{tabular} 
\\
$B_{RBp}$                                & 36                            & Bytes per RB pair (QPSK modulation)                                                                                                                                                            \\
$B_{req}$                                & 7                             & RRC Request message size (bytes)                                                                                                                                                                    \\
$B_{comp}$                               & 20                            & RRC Setup Complete message size (bytes)                                                                                                                                                             \\
$B_{s-comp}$                             & 13                            & RRC Security Mode Complete size (bytes)                                                                                                                                                                \\
$B_{r-UL}$                               & 10                            & RRC Reconfiguration Complete size (bytes)                                                                                                                                                   \\
$B_{data}$                               & 100                           & Data payload size (bytes)                                                                                                                                                                                      \\
$B_{comp_{CP}}$                            & 129                           & \begin{tabular}[c|]{@{}l@{}}RRC Setup Complete + NAS control plane \\ SR + $B_{data}$ message size for CP (bytes) \end{tabular}                                                            \\
$B_{data_{CP}}$                            & 120                           & \begin{tabular}[c|]{@{}l@{}}RRC UL inf. transfer + NAS control plane \\ SR + $B_{data}$ message size for CP (bytes) \end{tabular}                                                                 \\\hline
\end{tabular}

\end{table}

\section{Experimental setup and main results}
\label{sec:numresults}

We evaluate the energy consumption per packet of the three control procedures explained. For the radio resources available in the evaluation, we assume: system bandwidth of 5MHz (25 RBs), one RA opportunity every 5ms, 3 symbols for PDCCH per subframe, format 1 of the PDCCH, a fragmentation threshold of 6 RB pair, 54 available preambles, and a backoff window $W_{c}$ of 20 ms. We assume the path loss model of \cite{3gpp45820}, and a detection probability of 1. We consider an IAT which ranges from 320 ms to 48 hours. Additionally, the maximum allowed number of RA retries $m$ is 9 \cite{Madueño16}. Table \ref{table:var} summarizes other parameters' values considered in the evaluation. We consider an example of the parameters' usual value ranges, although the values are available from different sources.

\begin{figure}[!bt] 
	\centering
    \includegraphics[width=\columnwidth,height=5cm]{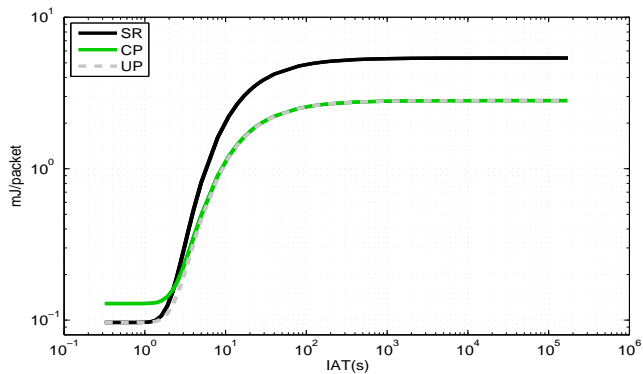}
	\caption{Average energy consumption per packet considering only the \textit{Communication} operation mode for SR, CP and UP ($P_{out}=0$).}
	\label{fig:EcomparisonSIG}
\end{figure}

Figure \ref{fig:EcomparisonSIG} shows the average energy consumption per packet only considering \textit{Communication} operation mode. As expected, CP and UP optimizations reduce the energy consumption compared to SR. However, for small IATs the CP solution is worst. This is due to data packets are sent as signaling. 

\begin{figure}[!bt] 
	\centering
    \includegraphics[width=\columnwidth,height=5cm]{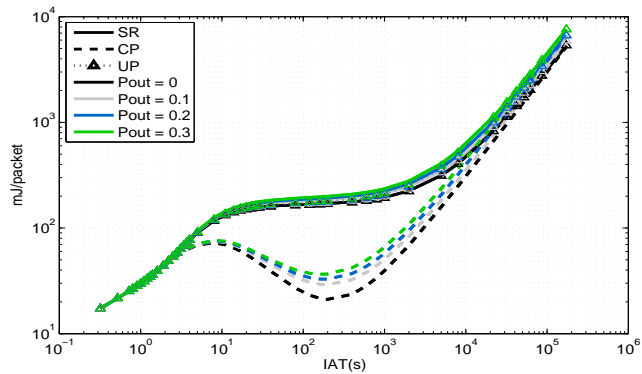}
	\caption{Average energy consumption per packet for SR, CP and UP, and different $P_{out}$.}
	\label{fig:EcomparisonPout}
\end{figure}

Figure \ref{fig:EcomparisonPout} shows the average energy consumption per packet for different IATs when the outage probability $P_{out}$ increases. The results include the three devices' operation modes, as described in Subsection \ref{subsec:transmod}. For small IATs, the packet transmissions while the device is in \texttt{RRC Connected} dominates the energy consumption. As the transmissions' frequency is reduced, \texttt{Connected Mode DRX} and PSM energy consumption begin to prevail. 
For both large and small IATs, the three procedures achieve similar results. However, 
for medium IATs the CP procedure reduces up to 87\% the energy consumption per packet. 
This is due to the reduced period of time in \textit{Inactive} state after the last transmission. However, CP procedure is the one most affected by an increase of $P_{out}$. As the energy consumption during signaling states and \textit{Inactive} state are similar in CP, the retransmissions to connect to the network 
impact CP results.

\begin{figure}[!bt] 
	\centering
  \includegraphics[width=0.9\columnwidth,height=4.3cm]{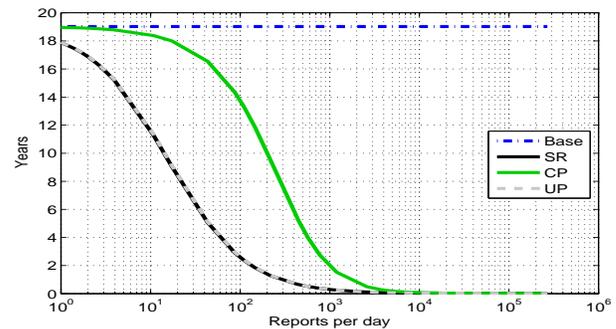}
	\caption{Battery Lifetime Estimation for SR, CP and UP procedures.}
	\label{fig:batlife}
\end{figure}

Further, from the average energy consumption per packet, in Figure \ref{fig:batlife} we estimate the device's battery lifetime with a battery capacity of 5Wh, as \cite{3gpp45820}. Battery lifetime's upper bound (labeled as Base) represents a device always in PSM. CP shows an extended battery lifetime for almost all values.

\section{Conclusion}
\label{sec:conclusion}

In this paper, we analyze and compare the average energy consumption of an IoT device when it sends data packets in LTE. The data transmission procedures analyzed are: conventional Service Request (SR), Control Plane (CP), and User Plane (UP) optimizations. Particularly, we have presented a Markov chain model to estimate the average energy consumption per packet. Our analysis includes different device's operation modes: \textit{Off}, \textit{Communication}, and \textit{Inactive}. 

The conducted evaluation highlights there is not an optimal solution which fits all data rates. For medium and big IATs, the energy consumption of \texttt{Connected Mode DRX} and PSM dominates the results. Within the medium IATs, CP optimization reduces up to 87\% the energy consumption for the considered scenario. This reduction is due to CP's Release Assistance Indication. This indication informs the MME whether can immediately release the connection. Despite we consider an example scenario, the CP's release optimization would continue to reduce the energy consumption compared to UP and SR for other scenarios. Finally, for large and small IATs, the results of the three procedures are similar.

\section*{Acknowledgment}
This work is partially supported by the Spanish Ministry of Economy and Competitiveness and the European Regional Development Fund (Projects TIN2013-46223-P, and TEC2016-76795-C6-4-R), and the Spanish Ministry of Education, Culture and Sport (FPU Grant 13/04833).
\bibliographystyle{IEEEtran}
\bibliography{refsgccicc}

\begin{thebibliography}{10}
\providecommand{\url}[1]{#1}
\csname url@samestyle\endcsname
\providecommand{\newblock}{\relax}
\providecommand{\bibinfo}[2]{#2}
\providecommand{\BIBentrySTDinterwordspacing}{\spaceskip=0pt\relax}
\providecommand{\BIBentryALTinterwordstretchfactor}{4}
\providecommand{\BIBentryALTinterwordspacing}{\spaceskip=\fontdimen2\font plus
\BIBentryALTinterwordstretchfactor\fontdimen3\font minus
  \fontdimen4\font\relax}
\providecommand{\BIBforeignlanguage}[2]{{%
\expandafter\ifx\csname l@#1\endcsname\relax
\typeout{** WARNING: IEEEtran.bst: No hyphenation pattern has been}%
\typeout{** loaded for the language `#1'. Using the pattern for}%
\typeout{** the default language instead.}%
\else
\language=\csname l@#1\endcsname
\fi
#2}}
\providecommand{\BIBdecl}{\relax}
\BIBdecl

\bibitem{Hoymann16}
C.~Hoymann, D.~Astely, M.~Stattin, G.~Wikstrom, J.~F. Cheng, A.~Hoglund,
  M.~Frenne, R.~Blasco, J.~Huschke, and F.~Gunnarsson, ``{{LTE release 14
  outlook}},'' vol.~54, no.~6, pp. 44--49, 2016.

\bibitem{Wang16}
X.~Wang, M.~J. Sheng, Y.~Y. Lou, Y.~Y. Shih, and M.~Chiang, ``{{Internet of
  Things Session Management Over LTE -Balancing Signal Load, Power, and
  Delay}},'' \emph{IEEE Internet of Things Journal}, vol.~3, no.~3, pp.
  339--353, June 2016.

\bibitem{3gpp45820}
3GPP, ``{{TR 45.820 Cellular system support for ultra-low complexity and low
  throughput Internet of Things (CIoT)}},'' Rel 13 v13.1.0, 2015.

\bibitem{Madueño16}
G.~C. Madue{\~{n}}o, J.~J. Nielsen, D.~M. Kim, N.~K. Pratas, \v{C}
  Stefanovi\'{c}, and P.~Popovski, ``{{Assessment of LTE Wireless Access for
  Monitoring of Energy Distribution in the Smart Grid}},'' \emph{IEEE Journal
  on Selected Areas in Communications}, vol.~34, no.~3, pp. 675--688, 2016.

\bibitem{3gpp23401}
3GPP, ``{TS} 23.401 {GPRS} enhancements for {E}volved {U}niversal {T}errestrial
  {R}adio {A}ccess {N}etwork access,'' Rel 14 v14.1.0, 2016.

\bibitem{3gpp36321}
{{3GPP}}, ``{{TS 36.321 Evolved Universal Terrestrial Radio Access; Medium
  Access Control protocol specification}},'' Rel 14 v14.0.0, 2016.

\bibitem{3gpp23682}
3GPP, ``{{TS 23.682 Architecture enhancements to facilitate communications with
  packet data networks and applications}},'' Rel 14 v14.1.0, 2016.

\bibitem{Haro15}
C.~Anton-Haro and M.~Dohler, \emph{{Machine-to-machine Communications
  Architecture, Performance and Applications}}.\hskip 1em plus 0.5em minus
  0.4em\relax Elsevier, Jan. 2015.

\bibitem{3gpp36213}
3GPP, ``{{TS 36.213 Evolved Universal Terrestrial Radio Access (E-UTRA);
  Physical layer procedures}},'' Rel 14 v14.1.0, 2016.

\bibitem{3gpp36912}
{{3GPP}}, ``{{TR 36.912 LTE; Feasibility study for Further Advancements for
  E-UTRA (LTE-Advanced)}},'' Rel 13 v13.0.0, 2016.

\bibitem{3gpp37869}
{3GPP}, ``{{TR 37.869 Study on Enhancements to MTC and other Mobile Data
  Applications; RAN aspects}},'' Rel 12 v12.0.0, 2013.

\bibitem{webDRX}
\BIBentryALTinterwordspacing
M.~Sauter. (2016) {{LTE Air Interface DRX Settings in Practice}}. [Online].
  Available:
  \url{https://blog.wirelessmoves.com/2016/05/lte-air-interface-drx-settings-in-practice.html}
\BIBentrySTDinterwordspacing

\bibitem{3gpp083917}
{{3GPP}}, ``{{R2-083917 Delay analysis for idle to active transition}},'' 2008.

\end{thebibliography}

\end{document}